\documentstyle[twoside,fleqn,espcrc2,psfig]{article}

\newcommand{\AmS}{{\protect\the\textfont2
  A\kern-.1667em\lower.5ex\hbox{M}\kern-.125emS}}

\title{ {\vskip -3.8cm 
\hfill{\small hep-ph/9712454}\\
\vskip -0.2cm 
\hfill{\small FTUV/97-70}
\vskip -0.2cm 
\hfill{\small IFIC/97-101}
\vskip -0.25cm 
\hfill{\small FISIST/13-97/CFIF}}
\vskip 1.1cm
Bounds on sterile neutrino mixing for 
	cosmologically interesting mass range
\thanks{Talk given by H. Nunokawa 
in Taup97, Gran Sasso, Italy, 7-11 September, 1997}
}
\author{H. Nunokawa\address{Instituto de F\'{\i}sica Corpuscular - C.S.I.C., 
Departament de F\'{\i}sica Te\`orica, Universitat de Val\`encia\\
46100 Burjassot, Val\`encia, Spain},
J. T. Peltoniemi\address{Department of Physics, Box 9, 00014 
University of Helsinki, Finland},
A. Rossi\address{Dept. de Fisica, Inst. 
Superior Tecnico, 1096 Lisbon Codex, Portugal}
and  
J.W.F.Valle$^a$
}
       
\begin{document}

\begin{abstract}

This talk summarizes our recent work which studied 
the impact of resonant $\nu_e \rightarrow\nu_s$ and 
$\bar{\nu}_e\rightarrow\bar{\nu}_s$ ($\nu_s$ is a {\it sterile} 
neutrino) conversions on supernova physics, 
under the assumption that the mass of the sterile state 
is in the few eV -cosmologically significant range.

\end{abstract}

\maketitle

\section{Introduction}
It has been discussed that the resonant conversion (MSW effect) 
\cite{MSW} of electron neutrinos into some sterile state 
in the dense media of type-II supernova could lead to some 
nontrivial consequences \cite{MS1,Raffelt,SS,juha}. 
We have reanalysed the impact of such conversion on supernova 
physics assuming the mass of the sterile state to be in 
the cosmologically significant range, i.e. 1-100 eV, 
the range relevant as dark matter component in the universe
\cite{kolb}

In what follows we will consider the system of 
$\nu_e$ and $\nu_s$ (and their anti-partners ) 
with non-zero masses and mixings and neglect 
the mixing between other flavors.  
The mass spectrum of neutrinos with the sterile state, 
relevant for our discussion, could naturally 
appear in some models \cite{ptv}.

\section{The active-sterile neutrino resonant conversion}

The effective potential for the $\nu_e-\nu_s$ system is given by,  
\begin{equation}
\label{potential}
V_e  = \frac{\sqrt{2}G_F \rho}{m_N} (Y_e- \frac{1}{2}Y_n)=
\frac{\sqrt{2}G_F}{2m_N}  \rho(3Y_e- 1),
\end{equation}
where $G_F$ is the Fermi constant, $\rho$ is the matter density,  $m_N$ is 
the nucleon mass and  $Y_e$ and $Y_n = 1-Y_e$ are the net 
electron and neutron number per baryon, respectively. 
For anti-neutrino system $V_e$ should be replaced by $-V_e$.

The resonance condition is given as
\begin{equation}
V_e = \frac{\delta m^2}{2E_\nu} \cos2\theta ~,
\label{rc}
\end{equation}
where $\delta m^2$ is the mass squared difference, 
$E_\nu$ is the neutrino energy and $\theta$ is the 
mixing angle. We assume $\delta m^2 > 0$, i.e., 
the heavier state is mostly sterile state.  

In general, in the region above the neutrinosphere 
the density decreases as $r$ (the radial 
distance from the center) increases. 
On the other hand, the electron fraction $Y_e$ 
takes minimum value nearby the neutrinosphere 
due to the efficient neutronization process 
and then $Y_e$ increases as $r$ increases. 
	
In Fig. 1 (a) the behavior of the matter density $\rho$ and 
the electron fraction $Y_e$ above the neutrinosphere are 
schematically shown. 
In Fig. 1 (b) we plot the behavior of the potential which can be inferred
from the behaviors of $\rho$ and $Y_e$ in (a). 

From Fig. 1 (b) we notice that for $\delta m^2 >$ 0 
as neutrinos leave from the neutrinosphere to the outer region, 
first $\bar{\nu}_e$'s undergo resonant conversion 
and then $\nu_e$'s are converted. 
The latter could undergo the transition twice if 
the values of $(\delta m^2/2E)\cos2\theta$ is smaller than 
the maximum value of $V_e$ (see Fig. 1(b)). 
In this work we appropriately take into account the double 
resonances for the $\nu_e \rightarrow\nu_s$ channel. 
\begin{figure}[htb]
\centerline{\protect\hbox{
\psfig{file=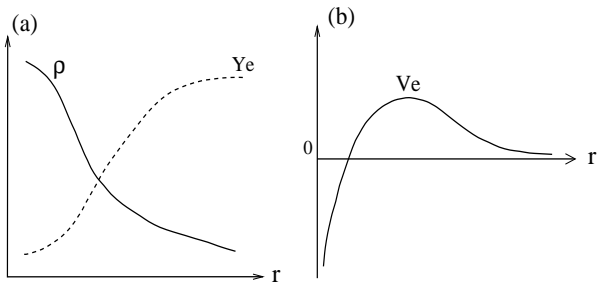,height=3.7cm,width=8.0cm,angle=-90}
}}
\vskip -0.7cm
\caption{In (a) the behavior of the density $\rho$ and $Y_e$
above neutrinosphere are schematically plotted and in (b) 
the potential $V_e  \propto \rho(3Y_e-1)$ is plotted. 
}
\vskip -0.5cm
\end{figure}
\section{Impact of the conversion on supernova physics}
In this section we discuss the effect of 
$\nu_e \rightarrow\nu_s$ and $\bar{\nu}_e\rightarrow\bar{\nu}_s$ 
conversions on shock re-heating, $\bar{\nu}_e$ signal
and heavy elements nucleosynthesis using the 
$\rho$ and $Y_e$ profiles from Wilson's supernova model. 

\subsection{Shock re-heating}

We first consider the neutrino-conversion effect on shock re-heating 
in the delayed explosion scenario \cite{delayed}. 
We estimate the neutrino energy deposition rate at 
the stalled shock with and without conversion and 
take the ratio which is defined to be $R$.  
It is clear that disappearance of either $\nu_e$ or $\bar{\nu}_e$ 
due to the resonant conversion into the sterile states 
will reduce the rate $R$. 
In Fig. 2 we plot the iso-contour for different values of 
the ratio $R$ in the parameter space ($\delta m^2, \sin^2 2 \theta$). 
We can conclude that if the neutrino re-heating is essential 
for successful supernova explosion the parameter region 
right to the curve, say $R=0.5$, is disfavoured. 
\begin{figure}[htb]
\vskip -0.8cm
\centerline{\protect\hbox{
\psfig{file=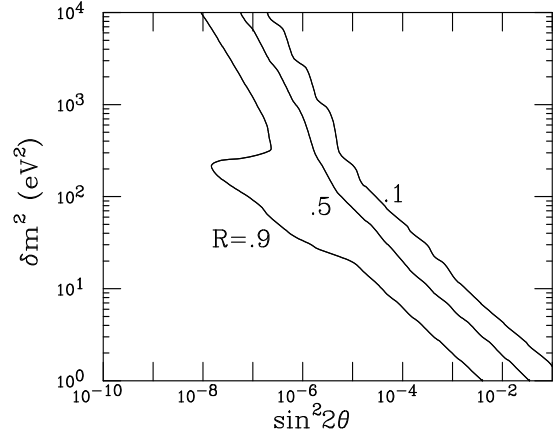,height=7.0cm,width=8.8cm,angle=90}
}}
\vglue -1.5cm
\caption{Contour plot of the ratio $R$ of the neutrino energy deposition 
behind the shock wave in the presence of conversions into 
sterile neutrinos, versus the case without conversions.
}
\label{fig:shockrate}
\vskip -0.5cm
\end{figure}

\subsection{SN1987A $\bar\nu_e$ signal}
Next we consider the impact on the observation of 
$\bar\nu_e$ signal on the earth. 
It is also clear that the resonant conversion of 
$\bar{\nu}_e\rightarrow\bar{\nu}_s$ could induce a reduction 
of $\bar\nu_e$ signal in the terrestrial detector. 
In Fig. 3 we present the contours of the 
$\bar\nu_e$ survival probability, properly averaged 
over the neutrino energy. We conclude that the successful 
observation of the $\bar\nu_e$ signal from supernova 
SN1987A in Kamiokande and IMB detectors \cite{kamimb} implies 
the absence of significant conversion 
of $\bar{\nu}_e\rightarrow\bar{\nu}_s$, 
disfavouring the parameter region right to the 
curve, say $P=0.5$. 
\begin{figure}[htb]
\vskip -0.8cm
\centerline{\protect\hbox{
\psfig{file=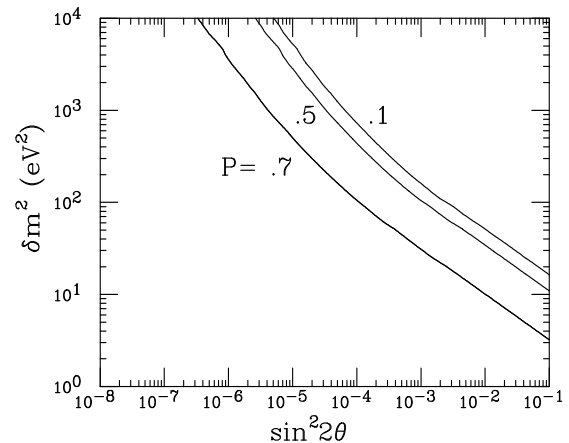,height=7.0cm,width=8.8cm,angle=90}
}}
\vglue -1.5cm
\caption{
Contour plots of the survival probability $P$ (figures at the curve) 
for the $\bar\nu_e\rightarrow\bar\nu_s$ conversion. 
}
\vskip -0.7cm
\end{figure}

\subsection{r-process}

Finally we discuss the impact of the neutrino conversion on 
heavy elements nucleosynthesis, so called $r$-process 
in supernova \cite{qian}. 
As discussed in ref. \cite{qian} one of the most 
relevant physical parameter in the $r$-process
is the electron fraction $Y_e$. To have successful
$r$-process the site must be neutron rich, i.e. 
$Y_e < 0.5$. 
The $Y_e$ value is mainly 
determined by the competition between the 
following two absorption reactions: 
\begin{equation}
\label{absorptions}
\begin{array}{@{\,}ll}
\nu_e+n      & \rightarrow  p+e^-, ~ \\
\bar\nu_e+p  & \rightarrow  n+e^+. 
\end{array}
\end{equation}
In the standard supernova model the latter process 
is favoured due to the higher average energy of 
$\bar\nu_e$ guaranteeing the neutron richness. 

We expect that $\nu_e \rightarrow\nu_s$ 
($\bar{\nu}_e\rightarrow\bar{\nu}_s$) 
conversion induce a decrease (increase) of $Y_e$ 
due to the decrease of the first (second) 
neutrino-absorption reaction in eq. (\ref{absorptions}).
The decrease of $Y_e$ implies that the site becomes more neutron 
rich and the $r$-process could be enhanced \cite{juha} whereas the 
increase of $Y_e$ induces the suppression of the $r$-process. 
Therefore, depending on which conversion channel is more 
efficient the $r$-process could either be enhanced or 
suppressed. 
In Fig. 4 we present the effect of neutrino conversion 
on the value of $Y_e$. In the region right to the curve 
$Y_e=0.5$ the value of $Y_e$ is larger than 0.5 and 
hence the $r$-process is suppressed. 
On the other hand, in the region delimited by the curve 
$Y_e=0.4$ the value of $Y_e$ could be decreased compared
to the standard case, leading to the enhancement of 
$r$-process. We note that due to non-trivial 
``feedback'' effect the value of $Y_e$ is not 
expected to be smaller than 1/3. See ref. \cite{sterile}
for more discussion.
\vskip -0.2cm
\begin{figure}[htb]
\vskip -0.7cm
\centerline{\protect\hbox{
\psfig{file=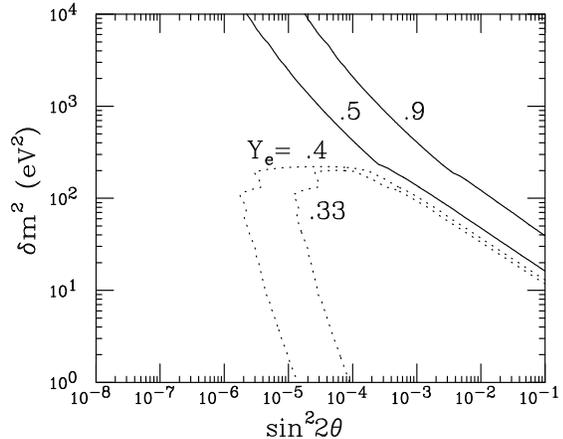,height=7.0cm,width=8.8cm,angle=90}
}}
\vglue -1.5cm
\caption{
Contour plots for the electron concentration $Y_e$ (figures at the curves)
in the region relevant for $r$-process. 
}
\vskip -0.7cm
\end{figure}
\section{Conclusion}
We have studied the impact of the resonant conversion of 
electron neutrinos into sterile state whose 
mass is assumed to be in the cosmologically interesting range. 
We have derived bounds on neutrino parameters from the shock 
re-heating, SN1987A $\bar{\nu}_e$ signal as well as 
$r$-process and we also found some parameter region 
where $r$-process could be enhanced. 
More detailed discussion on this work is found in ref. \cite{sterile}. 

\end{document}